\documentclass[letters,fleqn,usenatbib]{mnras}

\usepackage{newtxtext,newtxmath}
 
\usepackage[dvipsnames]{xcolor}

\usepackage[T1]{fontenc}
\usepackage{ae,aecompl}

\usepackage{graphicx}	
\usepackage{amsmath}	
\usepackage{amssymb}	

\def\gtorder{\mathrel{\raise.3ex\hbox{$>$}\mkern-14mu
     \lower0.6ex\hbox{$\sim$}}}
\def\ltorder{\mathrel{\raise.3ex\hbox{$<$}\mkern-14mu
     \lower0.6ex\hbox{$\sim$}}}

\title[Magnetic flux inversion]{Magnetic flux inversion in a peculiar changing look AGN}

\author[N. Scepi et al.]{
Nicolas Scepi$^{1}$\thanks{E-mail: nisc6580@colorado.edu}
Mitchell C. Begelman,$^{1,2}$
and Jason Dexter$^{1,2}$
\\
$^{1}$JILA, University of Colorado and National Institute of Standards and Technology, 440 UCB, Boulder, CO 80309-0440, USA \\
$^{2}$Department of Astrophysial and Planetary Sciences, University of Colorado, 391 UCB, Boulder, CO 80309-0391, USA}

\date{Accepted XXX. Received YYY; in original form ZZZ}

\pubyear{2020}

\begin{document}
\label{firstpage}
\pagerange{\pageref{firstpage}--\pageref{lastpage}}
\maketitle

\begin{abstract}
We argue that the changing-look event in the active galactic nucleus 1ES 1927+654, followed by a dip of 3 orders of magnitude in the X-ray luminosity, is controlled by a change in the accretion rate and an inversion of magnetic flux in a magnetically arrested disk (MAD). Before the changing-look event, strong magnetic flux on the black hole powers X-ray emission via the Blandford-Znajek process while the UV emission is produced by a radiatively inefficient magnetized disk. An advection event, bringing flux of the opposite polarity, propagates inward leading, first, to a rise in the UV/optical luminosity and, then, to a dip in the X-ray luminosity when it reaches the black hole. We estimate the timescale for magnetic flux advection and find that the observed timescale between the beginning of the changing-look event and the minimum in the X-ray luminosity, $\approx200$ days, is in agreement with the time needed to cancel the magnetic flux in a MAD extending to $\approx180\:r_g$.  Although flux inversion events might be rare due to the large ratio of flux-to-mass that is needed, we argue that AGN showing an unusually high ratio of X-ray to UV luminosity are prime candidates for such events. We also suggest that similar events may lead to jet interruptions in radio-loud objects.  
\end{abstract}




\section{Introduction}
Active galactic nuclei (AGN) are highly variable objects in the optical/UV, typically  showing rms variability of 10-20\% and occasionally changes by a factor as large as 2 in their optical luminosity \citep{rumbaugh2018}. The most extreme variability is found in ``changing-look'' AGN, which are observed to transit from Type 1 to Type 2 or vice-versa on a typical timescale of a few months. Type 1 AGN exhibit both narrow and broad emission lines while Type 2 AGN exhibit only narrow emission lines. The change from Type 2 to Type 1 is generally accompanied by a rise in the optical luminosity by a factor of $\simeq 2-10$ (or a corresponding decrease for Type 1 to Type 2). While the first changing-look AGN were detected in nearby Seyferts (\citealt{tohline1976}, \citealt{cohen1986}, \citealt{storchi1995}), they are now also detected in luminous quasars at higher redshifts (\citealt{lamassa2015}, \citealt{ruan2016}, \citealt{runnoe2016}, \citealt{macleod2016}, \citealt{yang2018}, \citealt{wang2018}).

The difference between Type 1 and Type 2 AGN has long been attributed to obscuration along the line of sight to the central region by a dusty torus \citep{antonucci1993}. It seems unlikely that a dusty cloud appearing or disappearing from the line of sight could be the driver of changing-look events \citep{trakhtenbrot2019} since it would need to cover a large fraction of the broad line region and move very fast to explain the timescales observed. A change in the intrinsic accretion power of AGN seems a more plausible explanation for changing-look events. This idea is supported by a few changing-look AGN that transited to or from a so-called ``true" Type 2 AGN, i.e., a Type 2 AGN showing no sign of obscuration (\citealt{lamassa2015}, \citealt{husemann2016}, \citealt{trakhtenbrot2019}).

One of these ``true'' Type 2 AGN, 1ES 1927+654 \citep{gallo2013}, recently underwent a peculiar changing-look event \citep{trakhtenbrot2019} that differentiates it from other changing-look AGN. Even before the changing-look event, 1ES 1927+654 had an X-ray luminosity in the 0.5-10 keV band approximately $1-2$ orders of magnitude higher than the bolometric luminosity inferred from the optical or UV \citep{ricci2020}, whereas AGN usually have an X-ray luminosity of the order of or lower than the UV luminosity \citep{svoboda2017}. The changing-look event, in which the optical luminosity rose by a factor of $\gtorder{10-100}$ and broad lines appeared, was then followed, near the peak of the UV luminosity, by a dramatic decrease of the X-ray luminosity by three orders of magnitude \citep{ricci2020}. This dip in the X-ray luminosity lasted $\approx 100$ days, after which the X-ray luminosity increased to a level ten times higher than before the event. Moreover, the UV/optical luminosity steadily decreased during the dip in the X-rays, suggesting that the optical changing-look event and the X-ray luminosity dip are unrelated. Usually in changing-look AGN, the X-ray luminosity follows the trend of the optical luminosity \citep{lamassa2015,husemann2016,parker2018,zetzl2018}.

\cite{ricci2020} suggested that the optical changing-look event and dip in the X-ray luminosity might be explained by the disruption of the inner accretion disk by a tidal disruption event (TDE). A TDE was suggested by a $t^{-5/3}$ fit to the rate of decrease of the UV luminosity.  However, \cite{trakhtenbrot2019} note that, due to  uncertainty in the time of the UV luminosity peak, the slope is poorly constrained. Moreover, emission lines usually seen in TDEs are absent in this case \citep{trakhtenbrot2019}. The smooth decrease of the UV luminosity during the X-ray luminosity dip also suggests that nothing as dramatic as the destruction of the inner disk happened after the optical changing-look event.

In this Letter, we propose an alternative scenario where the full series of events in 1ES 1927+654 is controlled by two main parameters: the rapid evolution of the magnetic flux close to the black hole (BH), and a change in accretion rate. While the timescales are much shorter than expected for viscous inflow in a geometrically thin accretion disk, they would be comparable to what one could expect in strongly magnetized elevated disks \citep{dexter2019}. In \S\ref{sec:before}, we argue that the high ratio of X-ray to UV luminosity in 1ES 1927+654 before the event could be explained by a magnetically arrested accretion disk (MAD). Then, in \S\ref{sec:inversion}, we show that the advection of matter threaded by magnetic flux of opposite polarity can rapidly destroy the MAD, leading first to a rise in the optical/UV luminosity and then a dip in the X-ray luminosity. A similar mechanism has been suggested to explain state transitions in X-ray binaries \citep{igumenshchev2009,dexter2014}. We estimate  
the amount of flux needed and the timescale on which it can be advected across the optical/UV-emitting region of the accretion disk and onto the black hole. The timescale is consistent with the observed decrease of the X-ray luminosity. In \S\ref{sec:source} we discuss the source of opposite polarity magnetic flux. We conclude in \S\ref{sec:conclusion} and discuss the prospects for observing similar phenomena in other sources. 

\section{Accretion state before and after the changing-look event}\label{sec:before}

\subsection{Compact X-ray corona }\label{sec:analytical_model}
Timing and spectral properties of hard X-ray emitting BHs are well described by an extremely compact X-ray emitter located within the few inner gravitational radii \citep{sanfrutos2013,reis2013,uttley2014}, consistent with the compact X-ray sizes inferred from quasar microlensing \citep{morgan2008,chartas2009,dai2009}. We assume that net magnetic flux deposited on the rotating BH extracts spin energy through the Blandford-Znajek (BZ) effect \citep{blandford1977}, and that a fixed fraction of the BZ power is radiated in  X-rays: $L_X=\epsilon_X P_\mathrm{BZ}$ where $\epsilon_X \equiv 0.1 \epsilon_{X,-1}$.  We adopt a modest reference value of $\epsilon_X$  because of the other channels available to the BZ power, such as a magnetized wind, but note that the efficiency could be higher \citep{crinquand2020}. The BZ power, in a split-monopole geometry, can be expressed as 
\begin{equation}\label{eq:PBZ}
P_\mathrm{BZ}\approx6\times10^{43} F_a M_7^{-2}\Phi_{X,30}^2\:\mathrm{erg\:s^{-1}}
\end{equation}
\citep{blandford1977,tchekhovskoy2010,sikora2013}, where $F_a=x_a^2f(x_a)$, $x_a=0.5a(1+\sqrt{1-a^2})^{-1}$, $f(x_a)\approx1+1.38x_a^2-9.2x_a^4$, $a$ is the dimensionless spin of the BH,  $\Phi_{X} \equiv 10^{30} \Phi_{X,30} \:\mathrm{G \:cm^2}$ is the amount of magnetic flux on the BH needed to power the X-rays,  and $M_\mathrm{BH} \equiv 10^7 M_7 M_\odot$ is the BH mass. We can rewrite the magnetic flux on the BH as 
\begin{equation}\label{eq:PhiX}
\Phi_X\approx 4\times 10^{30} \left(\frac{L_{X,43}}{\epsilon_{X, -1} F_{a,-1}} \right)^{1/2}M_7\:\mathrm{G\:cm^{2}},
\end{equation}
where $L_X \equiv 10^{43} L_{X,43}\:\mathrm{erg\:s^{-1}}$ is the X-ray luminosity and we have normalized $F_a$ to 0.1, its value for a BH with $a \approx 0.9$ (the limiting value for $a = 1$ is 0.2). Following the dip, the X-ray luminosity recovers to a level roughly 10 times that before the event, suggesting that the magnitude of $\Phi_X$ has increased by a factor of $\approx 3$.

\subsection{Accretion rate, radiative efficiency and magnetically arrested disk}
Magnetically arrested disks are the result of the accumulation, until saturation, of magnetic flux on a black hole by the ram pressure of the accretion flow \citep{narayan2003}. GRMHD simulations show that once the saturation point is reached, MADs alternate between episodes of expulsion of highly magnetized, low density fluid from the BH and advection of poloidal flux onto the BH \citep{igumenshchev2008}. The expulsions are due to non-axisymmetric instabilities such as the magnetic Rayleigh-Taylor instability. 

Once the MAD state is reached, any incoming flux in the disk will either reconnect with the background field (if of opposite polarity) or accumulate in the disk instead of advecting onto the BH (if of the same polarity). In the case of accumulation, we expect a strongly magnetized disk threaded by net magnetic flux to form. MADs thus provide the optimal conditions for powering a compact X-ray corona or jet, for a given accretion rate. 

To maintain the observed X-ray luminosity, the accretion rate must satisfy
\begin{equation}\label{eq:Mdot_MAD}
    {\dot{M}} \gtorder \dot{M}_\mathrm{MAD}\approx 2\times10^{-2}\left(\frac{L_{X,43}}{\epsilon_{X,-1}F_{a, -1}}\right)\:\mathrm{M_\odot\:yr^{-1}}
\end{equation}
\citep{mckinney2012}. We can compare the MAD limit to the accretion rate needed to produce the bolometric disk luminosity $L_d$, assuming a disk accretion efficiency of $\epsilon_d \equiv 0.1 \epsilon_{d,-1}$:  
\begin{equation}\label{eq:Mdot_MADratio}
    \frac{\dot{M}}{\dot{M}_\mathrm{MAD}}\approx \frac{\epsilon_X}{\epsilon_{d}} \frac{L_d}{L_X} F_{a, -1} \gtorder 1 .
\end{equation}
Before the changing-look event, $L_d/L_X \ltorder 0.1$ \citep{trakhtenbrot2019}.  This implies that the ratio of efficiencies must satisfy $\epsilon_X/\epsilon_d \gtorder 10$, suggesting that the UV/optical-emitting accretion disk is radiatively inefficient prior to the event.     

We can check the plausibility of this deduction by comparing $\dot M$ to the Eddington mass accretion rate, $10 L_\mathrm{Edd}/c^2$:
\begin{equation}\label{eq:Mdot_X}
    \frac{\dot{M}}{\dot{M}_\mathrm{Edd}}\approx 7\times10^{-3}\epsilon_{d, -1}^{-1}L_{d, 43}M_7^{-1}.
\end{equation}
For $\epsilon_d \approx 10^{-2}$ and $L_{d,43} \approx 0.1$ prior to the event, we estimate an Eddington factor that is marginally within the range that permits inefficient cooling in the inner disk \citep{esin1997} and that is consistent with the value of $\epsilon_d$ we used according to \cite{xie2019}.  If $L_d$ were much lower than $0.1 L_X$ before the changing-look event (which is possible because most of the pre-event optical data compiled by \cite{trakhtenbrot2019} are upper limits), this would expand the parameter space open to radiatively inefficient accretion, but at the expense of driving $\epsilon_X$ to values close to or exceeding one.

We therefore suggest that condition (\ref{eq:Mdot_MADratio}) is close to being marginally satisfied, i.e., that the disk prior to the changing-look event could be in a MAD state, $\dot M \approx \dot M_\mathrm{MAD}$.  A MAD, threaded by strong magnetic flux, could be radiatively inefficient regardless of the accretion rate. Strong, large-scale magnetic fields are very efficient at extracting angular momentum and energy in the form of magnetized outflows, while not depositing gravitational energy locally as would be the case with turbulent accretion \citep{ferreira1995}. Such fields are also able to support the disk vertically and decrease the density of the disk \citep{mishra2019}. These two effects could lead to MAD disks producing little UV/optical emission compared to the X-ray emission coming from the compact corona. Indeed, radiative simulations by \cite{morales2018} indicate a MAD radiative efficiency about 3 times lower than that of a standard thin disk \citep{novikov73}.

If the MAD state is to cause a low UV/optical radiative efficiency, it has to extend at least to the radius where a thin disk would peak in the optical, $r_\mathrm{opt}$. Estimating  $r_\mathrm{opt}$ using a local black-body model, we find
\begin{equation}
    r_\mathrm{opt}\approx1.7\times10^{2}\left(\frac{L_{X,43}}{\epsilon_{X,-1} F_{a,-1}}\right)^{1/3}M_7^{-2/3}r_g ,
\end{equation}
where $r_g = GM_{\mathrm BH}/c^2$. The magnetic flux required  to fill the MAD disk up to $r_\mathrm{opt}$, $\Phi_\mathrm{MAD}$, is then larger than $\Phi_X$ (equation \ref{eq:PhiX}) by a factor that scales $\propto (r_\mathrm{opt}/r_g)^{3/4}$,
where we assumed that the poloidal magnetic field follows the self-similar scaling $r^{-5/4}$
\citep{blandford1982,sikora2013}, 
but possibly with a small numerical coefficient due to the contribution of a  large toroidal field in holding the poloidal flux onto the BH (Scepi et al., in preparation).

During the changing-look event, the disk bolometric luminosity increases by a factor of at least $\approx 100$ at its peak.  We suggest that the actual accretion rate increases by about a factor of 10, with the other factor of 10 accounted for by an increase of $\epsilon_d$ from $\sim 10^{-2}$ to the standard thin disk value.  Such an increase of radiative efficiency is consistent with an increase in the Eddington factor to $\sim 0.1$, and also with a temporary increase in $\dot M/\dot M_\mathrm{MAD}$.  Moreover, a factor $\sim 10$ increase in $\dot M$ is also consistent with the long-term stable value of $L_X$ following its recovery from the dip, if a new MAD state is established close to the BH.  

In contrast, $L_d$ gradually declines following its peak.  We attribute this  either to the reestablishment of MAD conditions and an associated decrease of radiative efficiency sufficiently far out in the disk, or to a decrease in the mass accretion rate in the disk that has not yet had time to reach the BH. In the latter case, $L_X$ should eventually go down when the decrease in the accretion rate reaches the black hole.

\section{Mass accretion rate change and magnetic flux inversion as a transition mechanism}\label{sec:inversion}
\subsection{Magnetic flux inversion}\label{sec:Xray_change}

\begin{figure}
\includegraphics[width=80mm,height=100mm]{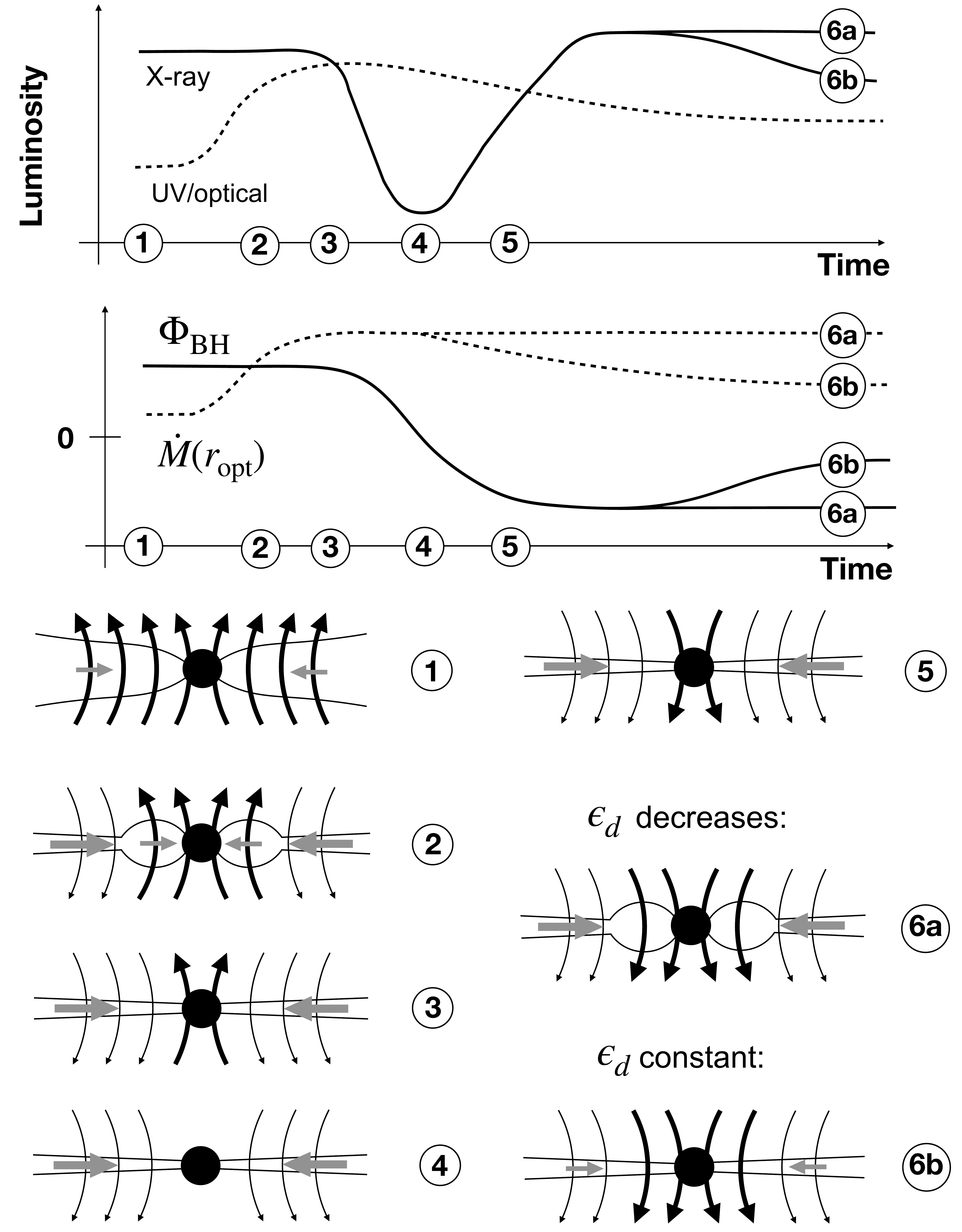}
\caption{Sketch of the sequence of events during the cancellation and reestablishment of magnetic flux in a MAD disk as well as the expected observational behavior.  The direction and thickness of the black, vertical arrows represent the sign and strength of the magnetic field, respectively. The length and thickness of the grey horizontal arrows represent the amplitude of the mass accretion rate. Geometrically thick/thin disks represent radiatively inefficient/efficient accretion flows. In the upper panels we also indicate inferred changes in the accretion rate during the changing-look transition of 1ES 1927+654. Curves 6a and 6b represent two possible scenarios for the future evolution of the disk.}
\label{fig:schema}
\end{figure}
On its own, a sudden change in the mass accretion rate is not sufficient to explain the dramatic reduction in the X-ray luminosity by a factor of 1,000, and its equally dramatic recovery. We suggest that this change is due to the annihilation and eventual replacement of the flux threading the BH by a sudden advection event bringing in magnetic field of opposite polarity. MADs are filled with a large poloidal magnetic field of one dominant polarity. By advecting enough  magnetic flux of the opposite polarity, one can destroy and then reestablish a MAD state \citep{mckinney2012,dexter2014}. 

The advected magnetic field will first reconnect in the outer strongly magnetized regions of the disk, leaving a disk with a lower magnetization and potentially higher radiative efficiency behind, increasing the UV/optical luminosity first. As the accretion event propagates inward, it will reach the BH and destroy the BZ corona. This leads to the dramatic decrease of $L_X$. The following increase of $L_X$ is due to the recreation of the corona by accumulation of the remaining net magnetic flux on the BH. Once the magnetic flux on the BH has reached saturation, the rest of the disk will continue accumulating net magnetic flux, reestablishing the MAD out to larger radii. Then, there are two options for explaining the decreasing UV luminosity. The onset of MAD may gradually reduce the disk radiative efficiency. Alternatively, the radiative efficiency might remain constant, indicating a decrease in the mass accretion rate. In the latter case, the X-ray luminosity will also decrease eventually. The sequence of events is sketched in Figure \ref{fig:schema} with a mock lightcurve as well as the physical states of the disk at different moments. This scenario is closely related to the magnetic flux paradigm of \cite{sikora2013} invoked to explain the radio-loud/radio quiet dichotomy in AGN.

In 1ES 1927+654, the recreation of the corona and the decrease in the UV luminosity are $\approx 4$ times slower than the destruction of the corona and the rise of the UV luminosity. This is consistent with the disk after the event being less magnetized and geometrically thinner than the original disk, increasing the timescale of advection of the flux as we will see in \S\ref{sec:timescales}. If the decrease of the UV luminosity after the event is due to a decrease in the accretion rate that has not yet reached the black hole, the X-ray luminosity decrease should occur on a timescale of $\approx 2$ years, $\approx4$ times the lag time between the beginning of the optical event and the destruction of the corona. 

\subsection{Magnetic field advection timescale}\label{sec:timescales}
We use two models to estimate the timescales for accretion and for advecting magnetic flux. First, we assume that in MADs, the accretion speed is dominated by the torque from a magnetized outflow \citep{ferreira1995}. If the magnetic flux is carried inward at roughly the same speed as the accreted mass \citep{scepi2020}, we obtain an advection speed for the magnetic flux of 
\begin{equation}
    v_\psi\sim \frac{4q}{\beta}\frac{H}{R}v_K ,
\end{equation}
where $q\equiv B_\phi/B_z=0.36\times\beta^{0.6}$ quantifies the strength of the magnetized outflow's torque \citep{scepi2020}, $\beta$ is the ratio of the thermal pressure to the magnetic pressure associated with the the $B_z$ component of the field, $H/R$ is the disk aspect ratio and $v_K\propto r^{-1/2}$ is the Keplerian velocity. The time it takes for an accretion event to go from $r_\mathrm{opt}$ to the event horizon of the black hole, $t_\mathrm{acc}\equiv\int_{r_g}^{r_\mathrm{opt}}dr/v_\psi$, is 
\begin{equation}
    \frac{t_\mathrm{acc}}{\mathrm{200\:days}}\approx10^{-3}\frac{R}{H}\left(\frac{q}{\beta}\right)^{-1}\left(\frac{L_{X,43}}{\epsilon_{X,-1} F_{a,-1}}\right)^{1/2}.
\end{equation}
For $\beta\approx10^{2}$, $L_{X,43}=10$, $\epsilon_{X,-1}=1$ and $H/R\approx0.3$, we have $t_\mathrm{acc}/200$ days $\approx10^{-1}$. 

If we assume instead that the accretion speed is given by a turbulent torque in a magnetically elevated disk \citep{dexter2019}, we find 
\begin{equation}\label{eq:alpha_acc}
    v_\psi\sim \alpha\left(\frac{H}{R}\right)^2v_K.
\end{equation}
This gives 
\begin{equation}
    \frac{t_\mathrm{acc}}{\mathrm{200\:days}}\approx4\times10^{-3}\alpha^{-1}\left(\frac{R}{H}\right)^{2}\left(\frac{L_{X,43}}{\epsilon_{X,-1} F_{a,-1}}\right)^{1/2}.
\end{equation}
Since the disk is magnetically supported, we have $H/R\approx0.3$. Assuming a reasonable $\beta\approx10^{2}$, we have $\alpha\approx1$ \citep{scepi2018b}. This also gives $t_\mathrm{acc}/200$ days $\approx10^{-1}$. In both scenarios the accretion event, coming from $\approx180\:r_g$, can explain the inversion event in the inferred time of 200 days for 1ES 1927+654.

\section{Source of magnetic flux}\label{sec:source}
The scenario that we propose requires a very high ratio of magnetic flux to advected mass,
\begin{equation}
\frac{\Phi_\mathrm{MAD}}{M_\mathrm{adv}}\approx3.7\times10^2\left(\frac{1}{\epsilon F_a}\frac{L_X}{10^{43}}\right)^{-1/4}M_7^{1/2}\mathrm{G\:cm^2\:g^{-1}}.
\end{equation}
 A self-gravitating molecular cloud typically has a magnetic flux to mass ratio of $10^{-4}-10^{-3}\:\mathrm{G\:cm^2\:g^{-1}}$ \citep{mouschovias1976}, which is $\approx$ 6 orders of magnitude below what we need to power the event. 
A tidal disruption event of a solar mass star threaded by a strong dipolar magnetic field of 1 MG would only a provide a magnetic flux to mass ratio of $\approx10^{-5}-10^{-4}\:\mathrm{G\:cm^2\:g^{-1}}$ and so could not power the event either.

Other sources of flux might include a hot ambient medium or streams of cool gas that are confined by  strong external thermal pressure rather than self-gravity, or an internal disk dynamo that provides the necessary inversion of the flux through stochastic processes. However, simple estimates suggest that the latter will develop rather slowly and thus may have a hard time leading to such rapid transitions \citep{begelman2014}. 

While the above represent ab initio sources of frozen-in flux, an alternate resolution is that the flux has been concentrated in the inner parts of the disk over a long period of time compared to the duration of the changing-look event.  A natural way to accomplish this would be for MAD conditions to have existed out to radii $\gtorder r_{\mathrm{opt}}$ (and possibly to much larger radii) long before the event.  Under such conditions, a large amount of matter could have been accreted while leaving its magnetic flux behind, increasing the flux to mass ratio by a large factor.  

The flux polarity reversal thus could have been ``baked" into the MAD long before the changing-look event.  Such a reversal would presumably evolve as a result of reconnection, but if it developed a smooth enough gradient it might be sufficiently long-lived to provide suitable pre-conditions for the changing-look transient.  We speculate that the sudden increase in $\dot M$ temporarily lifts the MAD condition since $\dot M \gg \dot M_\mathrm{MAD}$.  This could quench the strong field diffusion that characterizes the MAD state and force a large amount of flux to be advected inward, driving a reconnection front toward the black hole.

\section{Discussion and Conclusions}\label{sec:conclusion}
We have argued that the sudden disappearance of the X-ray emitting corona in the changing-look AGN 1ES 1927+654 resulted from the sudden change of polarity of the magnetic flux  advected onto the black hole.  If the corona is powered by the BZ effect, which is proportional to the square of the flux threading the BH, the decrease and reversal of this flux can lead to a dramatic decrease of the coronal power by orders of magnitude. The corona will be restored once the magnitude of the flux saturates again through continued advection.

We also argued that the accretion event associated with the magnetic flux inversion can trigger the increase of UV/optical flux (associated with the classical ``changing-look'' event) that preceded the destruction of the corona. An increase of the mass accretion rate by a factor of 10, coupled with an increase of the accretion efficiency by a factor of 10, are consistent with the overall increase of both the X-ray luminosity (following recovery from the dip) and the optical/UV luminosity at its peak. We consider two scenarios, depending on the MAD radiative efficiency, to explain the smooth decrease of the UV luminosity after the event. If MADs are radiatively inefficient, the UV luminosity decrease could simply be due to the recreation of the MAD state and the X-ray luminosity would not be expected to change after the recreation of the corona, provided that $\dot M$ remains at its elevated value. If MAD are radiatively efficient disks, the decrease of the UV luminosity could be due to a decrease in the accretion rate that has not yet reached the BH; in this case we expect the X-ray luminosity to follow the trend of the UV luminosity with a delay of $\approx 2$ years. We estimate that plausible initial conditions can trigger the changing-look event and the destruction-recreation of the corona in $\approx 200$ days (the time between the beginning of the changing-look event and the dip in $L_X$) for a BH mass of $2\times10^7 M_\odot$, as observed in 1ES 1927+654, provided that $\gtorder 10\%$ of the power from the BZ effect goes into X-rays.

For this coronal efficiency, before the event the disk was accreting at $\approx 10^{-2}\dot M_\mathrm{Edd}$ with a radiative efficiency of $\approx 1\%$. This low accretion rate is consistent with the absence of visible broad emission lines, according to the disk-wind scenario \citep{elitzur2016}. After the event the disk was accreting at $\approx 0.1 \dot  M_\mathrm{Edd}$ with a standard radiative efficiency of $\approx 10\%$. The increase in the accretion rate and the radiative efficiency can account for the sudden illumination of an existing broad line region (BLR) as suggested by the delay of $\approx 1-3$ months between the increase in the UV/optical luminosity and the appearance of the broad emission lines \citep{trakhtenbrot2019}. We note that a BLR at a few tens of light-days ($10^4\:r_g$ for 1ES 1927+654) \citep{trakhtenbrot2019} would be marginally consistent with an increase in the outflow mass loading rate up to $H/R=0.3$ in the 100 days of the changing-look event, though it would require an outflow velocity close to the local rotation velocity of the disk. Hence, the case of 1ES 1927+654 cannot settle whether the BLR was already sitting there or not when the change in the mass accretion rate and radiative efficiency illuminated it.

Our model is consistent with the idea that X-ray coronae in AGN can be extremely compact and associated with the BH itself rather than the accretion disk (\citealt{sanfrutos2013}, \citealt{reis2013}, \citealt{uttley2014}).  This model is especially consistent with the unusual AGN that exhibit a large X-ray-to-UV/optical luminosity ratio, since the disk reprocessing efficiency of X-rays from an ultra-compact corona will be small.  The accumulation of strong magnetic flux in the inner accretion disk can further accentuate the X-ray-to-UV/optical ratio. 

MAD simulations and the BZ effect are often associated with the creation of collimated jets, extending to large distances. Given the X-ray luminosity of 1ES 1927+654, and using equation (10) of \cite{sikora2013} to estimate the radio power of the jet,  one would expect 1ES 1927+654 to be radio-loud.  However, the radio-quiet nature of 1ES 1927+654 \citep{boller2003} suggests that the jet is somehow quenched, perhaps by interaction with the external medium, or internal instabilities leading to a lack of collimation. This is consistent with the coronal efficiency being large as the portion of BZ energy that does not end up as jet kinetic energy at large distances is potentially available for powering the X-rays.

We speculated on the origin of the magnetic flux necessary to trigger the inversion event and show that a very high flux to mass ratio is necessary. The ratio is much higher than what is expected from molecular clouds or a tidally disrupted star. If the flux is provided by an internal stochastic dynamo we would expect these events of full destruction of the corona to be rare because it would take a long time to develop such large flux inversions stochastically \citep{begelman2014}. Relatively long intervals between inversion events are also likely if the flux first has to accumulate out to large radii in the MAD.  In any case, these events should always be accompanied by a significant dip in $L_X$ that seems unrelated to the UV/optical light curve like in 1ES 1927+654. However, they could also happen in a different regime of accretion rate than in 1ES 1927+654 and the change in the UV/optical luminosity might be more moderate in certain cases. This emphasizes the need to monitor the X-ray as well as the optical in changing-look events.

Radio-loud objects, which are the AGN most likely to host a jet powered by the BZ effect and possibly a MAD, are prime candidates for undergoing similar events. These could exhibit a short-term interruption in the jet (possibly visible at milliarcsecond or higher resolutions) and a change in radio loudness associated mainly with an increase in the UV/optical emission. 

\section*{Acknowledgements}
We acknowledge financial support from  NASA Astrophysics Theory Program grants NNX16AI40G, NNX17AK55G, and 80NSSC20K0527 and an Alfred P. Sloan Research Fellowship (JD).





\bibliographystyle{mnras}
\bibliography{biblio} 


\bsp	
\label{lastpage}
\end{document}